\documentclass[aps,prl,twocolumn,showpacs,superscriptaddress]{revtex4}
\usepackage{amssymb}

\usepackage{graphicx}
\usepackage[english]{babel}
\usepackage{bm}    

\def\half{ {\frac{1}{2}} }

\def\reff#1{(\ref{#1})}
\newcommand{\be}{\begin{equation}}
\newcommand{\ee}{\end{equation}}
\newcommand{\<}{\langle}
\renewcommand{\>}{\rangle}
\renewcommand{\emptyset}{\varnothing}

\def\spose#1{\hbox to 0pt{#1\hss}}
\def\ltapprox{\mathrel{\spose{\lower 3pt\hbox{$\mathchar"218$}}
 \raise 2.0pt\hbox{$\mathchar"13C$}}}
\def\gtapprox{\mathrel{\spose{\lower 3pt\hbox{$\mathchar"218$}}
 \raise 2.0pt\hbox{$\mathchar"13E$}}}

\newcommand{\scrd}{{\cal D}}

\newcommand{\scrf}{{\cal F}}

\newcommand{\scrm}{{\cal M}}
\newcommand{\scrn}{{\cal N}}
\newcommand{\scro}{{\cal O}}

\newcommand{\scrs}{{\cal S}}

\newcommand{\scrz}{{\cal Z}}

\def\Z{{\mathbb Z}}

\begin{document}

\title{Dynamic critical behavior of the worm algorithm for the Ising model}

\author{Youjin Deng}
\affiliation{Department of Physics, New York University,
      4 Washington Place, New York, NY 10003, USA}
\author{Timothy M.~Garoni}
\affiliation{Department of Physics, New York University,
      4 Washington Place, New York, NY 10003, USA}
\author{Alan D. Sokal}
\affiliation{Department of Physics, New York University,
      4 Washington Place, New York, NY 10003, USA}
\affiliation{Department of Mathematics,
      University College London, London WC1E 6BT, UK}

\date{March 28, 2007; revised July 24, 2007}

\begin{abstract}
We study the dynamic critical behavior
of the worm algorithm for the two- and three-dimensional Ising models,
by Monte Carlo simulation.
The autocorrelation functions exhibit an unusual
three-time-scale behavior.
As a practical matter, the worm algorithm is slightly more efficient
than Swendsen--Wang for simulating the
two-point function of the three-dimensional Ising model.
\end{abstract}



\maketitle


Monte Carlo simulations in statistical mechanics \cite{Binder_79-92}
and quantum field theory \cite{Monte_Carlo_QFT}
typically suffer from {\em critical slowing-down}\/
\cite{Hohenberg_77,Sokal_Cargese_96}:
the autocorrelation (relaxation) time $\tau$ diverges
as the critical point is approached,
most often like $\tau \sim \xi^z$,
where $\xi$ is the spatial correlation length
and $z$ is a dynamic critical exponent.
For conventional local algorithms, one usually has $z \approx 2$.
This effect severely limits the efficiency of Monte Carlo studies
of critical phenomena in statistical mechanics
and of the continuum limit in quantum field theory.

One approach to circumventing this slowing-down involves
replacing the underlying spins or fields by an alternate representation,
obtained from the original model by algebraic transformation.
For instance, the celebrated
Swendsen--Wang (SW) cluster algorithm \cite{Swendsen_87}
simulates the $q$-state ferromagnetic Potts model \cite{Potts_52,Wu_82+84}
by passing back and forth between the Potts spin representation
and the Fortuin--Kasteleyn bond representation \cite{FK_69+72,Edwards_88}.
In this Letter we shall study another such algorithm,
namely the {\em worm algorithm}\/ \cite{Prokofev_01,Jerrum_93},
which simulates the high-temperature graphs of the spin model,
considered as a statistical-mechanical model in their own right.
Surprisingly, no systematic study of the dynamic critical behavior
of the worm algorithm has heretofore been carried out,
even in the simplest case of the Ising model.
As we shall show, the worm algorithm presents some unusual dynamical features,
which combine to make it an extraordinarily efficient algorithm for simulating
some (but not all) aspects of the three-dimensional Ising model.
Indeed, it is surprising (at least to us) that an algorithm
based on {\em local}\/ (``worm diffusion'') moves
could perform
so well.

We consider the zero-field ferromagnetic Ising model,
with nearest-neighbor coupling $J > 0$,
on a connected finite graph $G=(V,E)$ 
with vertex set $V$ and edge set $E$.
The high-temperature graphs of this model
are subsets $A \subseteq E$ of ``occupied bonds''.
For any bond configuration $A \subseteq E$,
we denote by $\partial A$
the set of vertices that touch an {\em odd}\/ number of bonds of $A$.
We focus attention on
the set $\scrs_\emptyset = \{A \colon\, \partial A = \emptyset\}$
of ``vacuum graphs''
and the sets $\scrs_{x,y} = \{A \colon\, \partial A = \{x,y\}\}$
of ``two-point-function graphs'',
with the convention that $\scrs_{x,x} = \scrs_\emptyset$.
Clearly, $\scrs_{x,y} = \scrs_{y,x}$.
The standard high-temperature expansion \cite{Thompson_79} states that
$Z = \sum_{A \in \scrs_\emptyset} w^{|A|}$ and
$Z G(x,y) = \sum_{A \in \scrs_{x,y}} w^{|A|}$,
where $Z$ is the Ising partition function (up to an uninteresting prefactor),
$G(x,y) = \< \sigma_x \sigma_y \>$
is the Ising two-point correlation function,
$w = \tanh J$,
and $|A|$ is the number of occupied bonds in the configuration $A$.

Instead of simulating the space of Ising spin configurations,
the worm algorithm simulates a space of high-temperature graphs.
More specifically,
the configuration space $\scrs$ of our version of the worm algorithm
consists of ordered triplets $(A,x,y)$ with $x,y \in V$
and $A \in \scrs_{x,y}$,
i.e., $A$ is a bond configuration having odd degree at $x$ and $y$
(unless $x=y$) and even degree at all other sites.
We set the weight of a configuration $(A,x,y)$ to be $w^{|A|} F(x,y)$
where $F$ is an arbitrary nonnegative function.
It follows that the ``partition function'' of our ensemble is
\begin{equation}
   \scrz \;=   \sum_{(A,x,y) \in \scrs} w^{|A|} F(x,y)
         \;=\; Z \sum_{x,y \in V} F(x,y) G(x,y)
         \;.
 \label{def.Zworm}
\end{equation}
We shall usually take $F \equiv 1$,
so that $\scrz = Z \< \scrm^2 \>$,
where $\scrm = \sum_x \sigma_x$ is the total magnetization;
in a translation-invariant situation $\scrz = ZV \chi$,
where $V$ is the volume and $\chi$ is the Ising susceptibility.
This configuration space is clearly tailored to studying
the Ising two-point function.

The elementary move of the worm algorithm is as follows:
Pick uniformly at random one of the two ``endpoints'' (say, $x$)
and one of the edges emanating from $x$ (say, $e = xx'$).
Propose to move from the current configuration $(A,x,y)$
to the new configuration $(A \triangle e,x',y)$,
where $\triangle$ denotes symmetric difference
(i.e., delete the bond $e$ if it is present, or insert it if it is absent).
Then accept or reject this move according to either the Metropolis
or the heat-bath criterion.
For instance, in the heat-bath version,
the configuration with (resp.\ without) $e$
gets probability $w/(1+w)$ [resp.\ $1/(1+w)$].
This simulates the distribution \reff{def.Zworm} with $F(x,y) = d_x d_y$,
where $d_x$ is the degree of the vertex $x$ in $G$
(i.e., the number of nearest neighbors).
Other choices of $F$ can be simulated by an appropriate
Metropolis accept-reject step.

Additional moves can optionally be added.
Because of the symmetry $x \leftrightarrow y$,
we can interchange $x$ and $y$ with probability 1/2 after each worm move.
More interestingly, whenever we reach $x=y$, we can move the endpoints
from $(x,x)$ to a randomly chosen $(x',x')$.
Finally, we can add ``local'' moves $A \mapsto A \triangle B$
(to be accepted or rejected according to the Metropolis criterion),
where $B$ is any bond configuration having $\partial B = \emptyset$,
e.g.\ a plaquette or a winding cycle.

We remark that the ``worm'' idea is very general:
enlarge a state space of ``vacuum'' (Eulerian) bond configurations
to include a pair of ``dislocations'', and then move those dislocations
by random walk.
This idea can be applied, in particular,
to the hexagonal-lattice $O(n)$ loop model \cite{ON_loop_papers}
at general $n$, 
and to vertex models.

In this Letter we report detailed measurements
of the dynamic critical behavior of the worm algorithm
and some of its variants,
for two- and three-dimensional Ising models at criticality,
on
lattices of size $L^d$ with periodic boundary conditions.
We shall measure time in units of ``hits'' of a single bond,
but we stress that the natural unit of time is one ``sweep'' of the lattice,
consisting of $L^d$ hits.
We write ${\bf z} = x-y$ for the vector distance
between the endpoints, and we define the observables
$\scrd_{{\bf a}} = \delta_{{\bf z},{\bf a}}$
and $\scrf_{{\bf p}} = e^{i {\bf p} \cdot {\bf z}}$.
In our simulations we measured
the number $\scrn = |A|$ of occupied bonds
as well as the short-distance observable
$\scrd_{{\bf 0}}$
and the low-momentum observable
$\scrf_{\rm low} = (1/2d) \sum_{|{\bf p}|=2\pi/L} \scrf_{{\bf p}}$.
Please note that $\< \scrd_{{\bf 0}} \> = 1/\chi$,
$\< \scrf_{\rm low} \> = \widetilde{G}({\bf p})/\chi$ for $|{\bf p}|=2\pi/L$,
and $\< \scrn \> = w \,\partial/\partial w \log(Z\chi)$.
In particular, the second-moment correlation length \cite{Salas-Sokal_2DIsing}
is $\xi = (\< \scrf_{\rm low} \>^{-1} - 1)^{1/2} / [2 \sin(\pi/L)]$.

For any observable $\scro$,
let $\rho_{\scro} (t)$ be its normalized autocorrelation function
in the stationary stochastic process.
Then define the exponential autocorrelation time
\begin{equation}
   \tau_{{\rm exp},\scro} \;=\;
   \limsup_{t \to \pm\infty} {|t| \over   - \log |\rho_{\scro}(t)|}
\end{equation}
and the integrated autocorrelation time
\begin{equation}
   \tau_{{\rm int},\scro}   \;=\;
   \half \sum_{t = -\infty}^{\infty}  \rho_{\scro}(t)
   \;.
\end{equation}
Typically all observables $\scro$
(except those that, for symmetry reasons,
 are ``orthogonal'' to the slowest mode)
have the same value $\tau_{{\rm exp},\scro} = \tau_{{\rm exp}}$.
However, they may have very different amplitudes of ``overlap''
with this slowest mode;
in particular, they may have very different values of the
integrated autocorrelation time,
which controls the efficiency of Monte Carlo simulations
\cite{Sokal_Cargese_96}.
We define dynamic critical exponents $z_{\rm exp}$
and $z_{{\rm int},\scro}$ by
$\tau_{\rm exp} \sim \xi^{z_{\rm exp}}$
and $\tau_{{\rm int},\scro} \sim \xi^{z_{{\rm int},\scro}}$,
where time is measured in ``sweeps''.
On a finite lattice at criticality, $\xi$ can here be replaced by $L$.

Before presenting our numerical results,
let us make some heuristic predictions for the dynamic behavior
of the worm algorithm.
Suppose first that the bond configuration $A$ is at all times
completely equilibrated for the given endpoints $x,y$ \cite{note_localmoves}.
Then ${\bf z} = x-y$ performs a random walk with drift
having equilibrium distribution $G({\bf z})/\chi$.
In the simplest case $G \equiv 1$
(which corresponds to the zero-temperature limit $J=\infty$),
the eigenvectors of this random walk are $\scrf_{{\bf p}}$,
with eigenvalues (in the heat-bath version)
$\lambda_{{\bf p}} = [1 + (1/d) \sum_{i=1}^d \cos p_i]/2$.
In particular, $\tau_{\rm exp} \sim L^2$.
Furthermore, the autocorrelation function of $\scrd_{{\bf a}}$
is ${\rm const} \times \sum_{{\bf p} \neq {\bf 0}} \lambda_{{\bf p}}^{|t|}$.
In the limit $L \to \infty$, this tends to $P_t({\bf 0})$,
the return probability of the corresponding random walk on $\Z^d$;
in particular, for large $t$ it behaves like $t^{-d/2}$.
It follows that $\tau_{{\rm int},\scrd_{{\bf a}}} \sim L^{2-d}$ in $d<2$,
$\log L$ in $d=2$, and $L^0$ in $d>2$.
We expect these scaling predictions to hold near the critical point in $d=1$
(since $J_c = \infty$)
and in the {\em low-temperature}\/ regime ($J > J_c$) in $d \ge 2$.
A more complicated behavior is likely to occur
in the critical regime in $d \ge 2$,
where $G({\bf z}) \sim |{\bf z}|^{-(d-2+\eta)}$ as $|{\bf z}| \to\infty$.
A Fokker--Planck analysis under the hypothesis of perfect equilibration
of bonds predicts \cite{worm_fullpaper}
$\rho_{\scrd_{{\bf a}}}(t) \sim t^{-(1-\eta/2)}$.


It should be noted, however, that
$\scrd_{{\bf 0}}$ estimates $\chi$ via a ``rare'' event,
i.e.\ a binomial random variable with probability $1/\chi$.
The variance of this random variable is also of order $1/\chi$,
so that $\sim \chi$ samples are needed to get a relative variance of order 1.
This is an example of ``statistical inefficiency due to
large static variance''.

Note, finally, that a simple variational argument
(following the model in \cite{Sokal_Cargese_96})
shows that $z_{{\rm int},\scrn} \ge \alpha/\nu$.

We began by simulating the worm algorithm at the critical point in $d=1$
and in the low-temperature phase in $d=2,3$.
The autocorrelation functions of $\scrf_{\rm low}$ and $\scrd_{\bf 0}$
behave exactly as predicted.
The autocorrelation function of $\scrn$ is essentially a pure exponential,
with autocorrelation time $\sim L^2$ in $d=1$,
$\sim L^2 \log L$ in $d=2$, and $\sim L^3$ in $d=3$.

We next simulated the square-lattice ($d=2$) Ising model
at criticality ($w_c = \sqrt{2} - 1$)
on lattices $4 \le L \le 2048$.
The simulation lengths varied from $5 \times 10^{11}$ hits ($L=4$)
to $5.4 \times 10^{14}$ hits ($L=2048$).
These runs used approximately 6.9 yr CPU time
on a 3.2 GHz Xeon EM64T processor.
Estimates of the static quantities $\chi$, $\xi$ and $E$
agreed within error bars with previous high-precision simulations
using the SW algorithm \cite{Salas-Sokal_2DIsing}.

\begin{figure}[t]
\begin{center}
\includegraphics[width=\columnwidth]{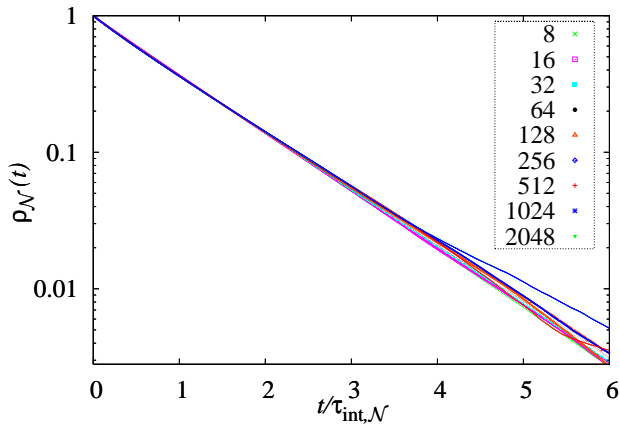}
\end{center}
\vspace*{-8mm}
\caption{
   Autocorrelation function $\rho_{\scrn}(t)$ versus $t/\tau_{{\rm int},\scrn}$
   for the critical two-dimensional Ising model.
}
\label{fig1}
\end{figure}

The slowest mode is the number $\scrn$ of occupied bonds,
and the decay of $\rho_{\scrn}(t)$ is very close to pure exponential
(Fig.~\ref{fig1}).
A fit $\tau_{{\rm int},\scrn}/L^2 \sim AL^z$ (resp.\ $AL^z + B$)
yields $z_{\rm exp} = z_{{\rm int},\scrn} \approx 0.379$ (resp.\ 0.338).
But a behavior $\tau_{{\rm int},\scrn}/L^2 \sim \log^2 L$
is also conceivable.

%
%

%
%



\begin{figure}[t]
\begin{center}
\includegraphics[width=\columnwidth]{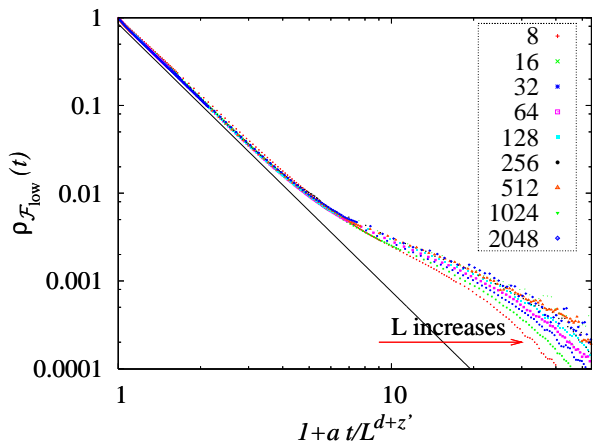}
\end{center}
\vspace*{-5mm}
\caption{
   Autocorrelation function $\rho_{\scrf_{\rm low}}(t)$
   versus $1 + at/L^{d+z'}$ with $a = 2.5$, $z' = 0.315$
   for the critical two-dimensional Ising model.
   Black line shows initial decay with power $r \approx 3.05$.
}
\label{fig3}
\end{figure}

A more complicated behavior is exhibited by $\scrf_{\rm low}$:
contrary to our simple-minded prediction,
its decay is far from a pure exponential.
Rather, it appears to scale like
$\rho_{\scrf_{\rm low}}(t) \approx f(t/L^{d+z'})$
with a scaling function $f$ that shows an initial power-law decay
$f(x) \sim x^{-r}$ with $r \approx 3.05$
but then bends towards an unknown smaller power (Fig.~\ref{fig3}).
We find $z' \approx 0.315$.
It is not clear whether $z'$ equals $z_{\rm exp}$ or is slightly smaller.

%
%

\begin{figure}[t]
\begin{center}
\includegraphics[width=\columnwidth]{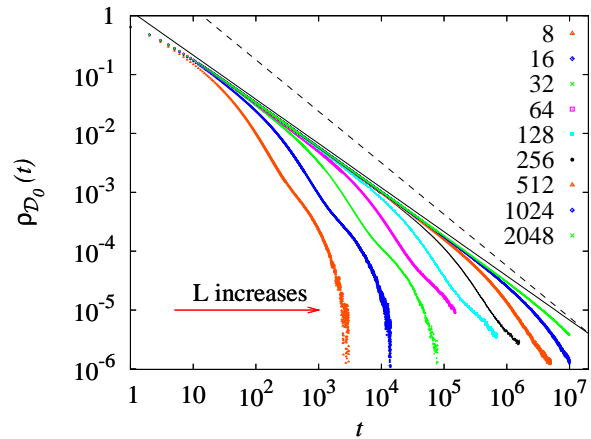}
\end{center}
\vspace*{-6mm}
\caption{
   Autocorrelation function $\rho_{\scrd_{{\bf 0}}}(t)$ versus $t$
   for the critical two-dimensional Ising model.
   Asymptotic straight line has power $s=0.75$.
   Dashed line has power $1-\eta/2 = 7/8$.
}
\label{fig4}
\end{figure}

The most interesting behavior of all is shown by $\scrd_{{\bf 0}}$,
which exhibits significant decorrelation on a time scale
of order 1 hits (Fig.~\ref{fig4}).
The data fall in the limit $L \to\infty$
on a beautiful scaling curve $\rho_{\scrd_{{\bf 0}}}(t) = g(t)$,
with $g(x) \sim x^{-s}$ as $x \to \infty$.
We find $s \approx 0.75$,
which is clearly smaller than our prediction $s=1-\eta/2 = 7/8$
based on perfect equilibration of bonds.
Apparently this latter prediction provides a 
{\em lower bound}\/ on $\rho_{\scrd_{{\bf 0}}}(t)$
and hence an {\em upper bound}\/ on the exponent $s$.

%
%

\begin{figure}[t]
\vspace*{-3mm}
\begin{center}
\includegraphics[width=\columnwidth]{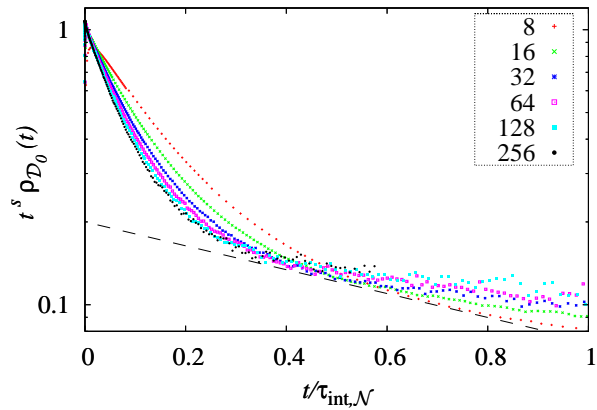}
\end{center}
\vspace*{-9mm}
\caption{
   $t^s \rho_{\scrd_{{\bf 0}}}(t)$ versus $t/\tau_{{\rm int},\scrn}$
   for the critical two-dimensional Ising model,
   with $s=0.75$.
   Dashed line shows the decay rate $\tau_{\rm exp}$.
}
\label{fig5}
\end{figure}

Finally, we analyzed the universal crossover from
short-time to long-time behavior in $\rho_{\scrd_{{\bf 0}}}$,
which we hypothesize is of the form
$\rho_{\scrd_{{\bf 0}}}(t) = g(t) \, h(t/L^{d+z_{\rm exp}})$,
by plotting $t^s \rho_{\scrd_{{\bf 0}}}(t)$ versus $t/\tau_{{\rm int},\scrn}$.
A fairly clear scaling curve is seen (Fig.~\ref{fig5}),
though it is noisy for large lattices.
Using this scaling Ansatz to compute the area under the curve of
$\rho_{\scrd_{{\bf 0}}}(t)$, we conclude that
\begin{equation}
   z_{{\rm int}, \scrd_{{\bf 0}}} \;=\;
      \cases{
          -sd + (1-s)z_{\rm exp}   &  if $s < 1$  \cr
          -d                       &  if $s > 1$
      }
 \label{eq.zintD0}
\end{equation}
Using $z_{\rm exp} \approx 0.338$ and $s \approx 0.75$,
we find $z_{{\rm int}, \scrd_{{\bf 0}}} \approx -1.42$.

%
%

We also studied the variant of worm algorithm
with the move $(x,x) \to (x',x')$.
We find that, in the limit $L \to\infty$,
this move has no effect
(i.e., the ratio of autocorrelation times with and without the move
 tends to 1).
Evidently the diffusion of the endpoints $x,y$
around the lattice in the basic worm algorithm is already sufficient.

We next simulated the simple-cubic ($d=3$) Ising model
at the estimated critical point $J_c = 0.22165455$ \cite{Deng_03}
on lattices $4 \le L \le 256$.
The simulation lengths varied from $5 \times 10^{10}$ hits ($L=4$)
to $7.2 \times 10^{14}$ hits ($L=256$).
These runs used approximately 6.6 yr CPU time.

\begin{figure}[t]
\begin{center}
\includegraphics[width=\columnwidth]{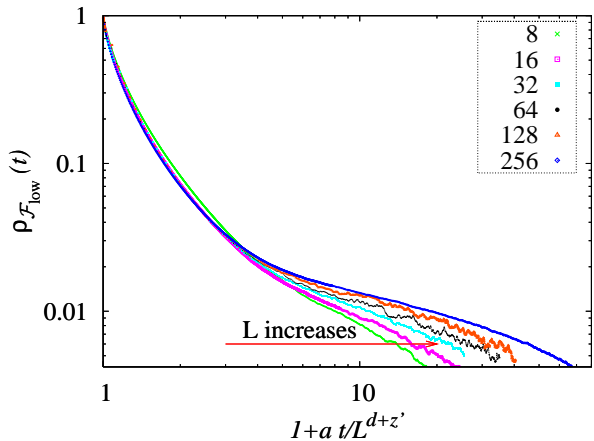}
\end{center}
\vspace*{-6mm}
\caption{
   Autocorrelation function $\rho_{\scrf_{\rm low}}(t)$
   versus $1 + at/L^{d+z'}$ with $a = 1.0$, $z' = -0.15$
   for the critical three-dimensional Ising model.
}
\label{fig6}
\end{figure}

\begin{figure}[t]
\begin{center}
\includegraphics[width=\columnwidth]{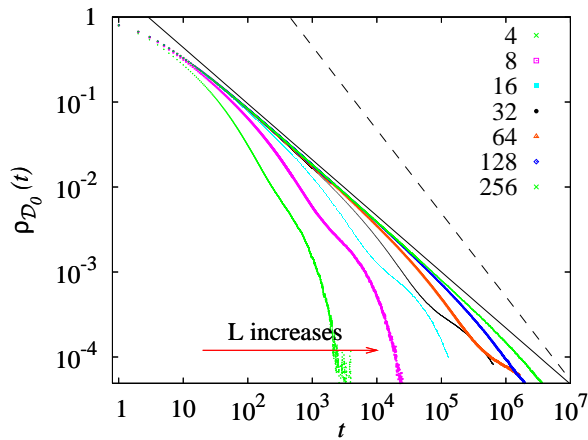}
\end{center}
\vspace*{-6mm}
\caption{
   Autocorrelation function $\rho_{\scrd_{{\bf 0}}}(t)$ versus $t$
   for the critical three-dimensional Ising model.
   Asymptotic straight line has power $s=0.66$.
   Dashed straight line has power $1-\eta/2 = 0.982$.
}
\label{fig7}
\end{figure}

Once again the slowest mode is $\scrn$,
and the decay is very close to pure exponential.
In contrast to $d=2$,
the exponent $z_{\rm exp} = z_{{\rm int},\scrn}$
now appears to {\em equal}\/ $\alpha/\nu \approx 0.174$.
Again $\scrf_{\rm low}$ exhibits a scaling curve
with a very strong bending and an unknown asymptotic decay exponent $r$,
with $z' \approx -0.15 < z_{\rm exp}$ (Fig.~\ref{fig6}).
Finally, $\scrd_{{\bf 0}}$ exhibits a clear scaling
with exponent $s \approx 0.66$ (Fig.~\ref{fig7}).
Again this exponent is smaller than our prediction
$s=1-\eta/2 \approx 0.982$ \cite{Deng_03}
based on perfect equilibration of bonds.
{}From \reff{eq.zintD0} we obtain
$z_{{\rm int}, \scrd_{{\bf 0}}} \approx -1.92$.

%
%

%
%

In summary, the worm dynamics for the critical Ising model
appears to exhibit decorrelation on three different time scales:
$\scrn$ has an almost-pure-exponential decay on the very long time scale
$L^{d+z_{\rm exp}}$ (in ``hits'');
$\scrf_{\rm low}$ has a complicated power-law decay on the long time scale
$L^{d+z'}$;
and $\scrd_{\bf 0}$ has a simple power-law decay on the short time scale $L^0$.
This is analogous to but more complex than the two-time-scale behavior
recently observed in the Sweeny dynamics for the random-cluster model
\cite{sweeny_prl}.

The practical efficiency of the worm algorithm depends on the observable.
For estimating $\chi$ using $\scrd_{\bf 0}$,
the autocorrelation time
$\tau_{{\rm int}, \scrd_{{\bf 0}}}
 \sim L^{z_{{\rm int}, \scrd_{{\bf 0}}}}$ [cf.\ \reff{eq.zintD0}]
must be multiplied by a factor $\chi \sim L^{\gamma/\nu}$
due to static variance, leading to an ``effective dynamic critical exponent''
$z_{{\rm eff},\scrd_{{\bf 0}}} =
 z_{{\rm int}, \scrd_{{\bf 0}}} + \gamma/\nu$,
i.e.\ $\approx 0.33$ in $d=2$ and $\approx 0.04$ in $d=3$.
This is slightly worse than the SW algorithm in $d=2$
but significantly better in $d=3$,
where $z_{\rm SW} \approx 0.46$ \cite{Ossola-Sokal}.
For estimating $\xi$ using $\scrf_{\rm low}$,
the exponent $z_{{\rm int}, \scrf_{\rm low}}$
is uncertain because of the uncertainty on the decay exponent $r$
(Figs.~\ref{fig3} and \ref{fig6}),
but it must lie somewhere between $z'$ and $z_{\rm exp}$,
i.e.\ from 0.31 to 0.38 in $d=2$
and from $-0.15$ to 0.17 in $d=3$.
This is again slightly worse than SW in $d=2$
but better in $d=3$.
We conclude that the worm algorithm is,
asymptotically as $L \to\infty$,
the most efficient algorithm currently available
for simulating $\chi$ and $\xi$ in the three-dimensional Ising model.
In practice, our data show \cite{worm_fullpaper}
that the worm algorithm outperforms SW when $L \gtapprox 32$,
at a rate that grows like $L^{\approx 0.32}$.


Details of these simulations and their data analysis
will be reported separately \cite{worm_fullpaper}.

\begin{acknowledgments}
We thank Fabien Alet and Nikolay Prokof'ev for helpful discussions.
This work was supported in part by NSF grants PHY--0116590 and PHY--0424082.
\end{acknowledgments}



\begin{thebibliography}{99}

\bibitem{Binder_79-92}  K. Binder, ed., {\em Monte Carlo Methods
   in Statistical Physics}\/, 2nd ed. (Springer-Verlag, Berlin, 1986).

\bibitem{Monte_Carlo_QFT}
    I. Montvay and G. M\"unster, {\em Quantum Fields on a Lattice}\/
    (Cambridge University Press, New York, 1994), chap.~7.

\bibitem{Hohenberg_77}
P.C. Hohenberg and B.I. Halperin, Rev. Mod. Phys. {\bf 49}, 435 (1977).

\bibitem{Sokal_Cargese_96}
A.D. Sokal, in {\em Functional Integration: Basics and Applications}\/,
   ed. C. de Witt-Morette, P. Cartier and A. Folacci
   (Plenum, New York, 1997), pp.~131--192.

\bibitem{Swendsen_87} R.H. Swendsen and J.-S. Wang,
   Phys. Rev. Lett. {\bf 58}, 86 (1987).

\bibitem{Potts_52}  R.B. Potts,
   Proc. Cambridge Philos. Soc. {\bf 48}, 106 (1952).

\bibitem{Wu_82+84}  F.Y. Wu, Rev. Mod. Phys. {\bf 54}, 235 (1982);
   {\bf 55}, 315 (E) (1983);  J. Appl. Phys. {\bf 55}, 2421 (1984).

\bibitem{FK_69+72}  P.W. Kasteleyn and C.M. Fortuin,
   J. Phys. Soc. Japan {\bf 26} (Suppl.), 11 (1969);
   C.M. Fortuin and P.W. Kasteleyn, Physica {\bf 57}, 536 (1972).

\bibitem{Edwards_88} R.G. Edwards and A.D. Sokal, Phys. Rev. D {\bf 38},
   2009 (1988).

\bibitem{Prokofev_01}  N. Prokof'ev and B. Svistunov,
   Phys. Rev. Lett. {\bf 87}, 160601 (2001).

\bibitem{Jerrum_93}  For a closely related algorithm,
   see M. Jerrum and A. Sinclair, SIAM J. Comput. {\bf 22}, 1087 (1993).


\bibitem{Thompson_79} C.J. Thompson, {\em Mathematical Statistical Mechanics}\/
   (Princeton Univ.\ Press, 1979), secs.~6--1 and 6--4.

\bibitem{ON_loop_papers}  E. Domany, D. Mukamel, B. Nienhuis
   and A. Schwimmer, Nucl. Phys. B {\bf 190}, 279 (1981);
   B. Nienhuis, Phys. Rev. Lett. {\bf 49}, 1062 (1982);
   J. Stat. Phys. {\bf 34}, 731 (1984).

\bibitem{Salas-Sokal_2DIsing}  J. Salas and A.D. Sokal,
   J. Stat. Phys. {\bf 98}, 551 (2000).

\bibitem{note_localmoves}  This could be achieved by interspersing worm moves
 with infinitely many local moves $A \mapsto A \triangle B$,
 where $B$ runs over a generating set of configurations
 having $\partial B = \emptyset$.

\bibitem{worm_fullpaper}  Y. Deng, T.M. Garoni and A.D. Sokal, in preparation.


\bibitem{Deng_03}  Y. Deng and H.W.J. Bl\"ote, Phys. Rev. E {\bf 68}, 036125
    (2003).

\bibitem{sweeny_prl}  Y. Deng, T.M. Garoni and A.D. Sokal,
   Phys. Rev. Lett. {\bf 98}, 230602 (2007), cond-mat/0701113.

\bibitem{Ossola-Sokal}  See G. Ossola and A.D. Sokal, Nucl. Phys. B {\bf 691},
   259 (2004), hep-lat/0402019
   for a summary of the latest data.







%
%
%
%
%
%
%
%
%
%
%
%


%
%

%





%
%
%
%
%
%
%
%
%
%

%
%
%
%
%



\end{thebibliography}
\end{document}